\documentstyle[ppmo209,twoside,epsf,fancyhdr]{article}
\input{psfig.sty}                               
\def\aj{{AJ}}

\def\annrev{{ARA\&A}}
\def\apj{{ApJ}}
\def\apjs{{ApJS}}
\def\asec{$^{\prime\prime}$}

\def\hbet{H$\beta$}

\def\kms{km s$^{-1}$}
\def\lamb{$\lambda$}
\def\lax{{$\mathrel{\hbox{\rlap{\hbox{\lower4pt\hbox{$\sim$}}}\hbox{$<$}}}$}}
\def\gax{{$\mathrel{\hbox{\rlap{\hbox{\lower4pt\hbox{$\sim$}}}\hbox{$>$}}}$}}
\def\simlt{\lower.5ex\hbox{$\; \buildrel < \over \sim \;$}}
\def\simgt{\lower.5ex\hbox{$\; \buildrel > \over \sim \;$}}
\def\lum{ergs s$^{-1}$}

\def\mnras{{MNRAS}}

\def\pasp{{PASP}}

\def\percm2{cm$^{-2}$}
\def\peryr{yr$^{-1}$}

\def\solmass{$M_\odot$}
\def\oii{[\ion{O}{2}]}

\def\hii{\ion{H}{2}}
\def\oiii{[\ion{O}{3}]}

\def\oi{[\ion{O}{1}]}

\def\hii{H~{\sc II}}
\def\oi{[O~{\sc I}]}
\def\oii{[O~{\sc II}]}
\def\oiii{[O~{\sc III}]}

\begin{document}

\markboth {Luis C. Ho} {AGNs and Starbursts: What Is the Real Connection?}

\title{AGNs and Starbursts: What Is the Real Connection?}

\author{Luis C. Ho$^{1}$}

\inst{$^1$ The Observatories of the Carnegie Institution of Washington, 
813 Santa Barbara St., Pasadena, CA 91101, USA }
\email{lho@ociw.edu}

\date{Received~~2005 November 7; accepted~~2005~~November 7}

\begin{abstract}
It is now widely believed that the growth of massive black holes is
closely linked to the formation of galaxies, but there have been few
concrete constraints on the actual physical processes responsible for
this coupling.  Investigating the connection between AGN and starburst
activity may offer some empirical guidance on this problem.  I summarize
previous observational searches for young stars in active galaxies, concluding
that there is now compelling evidence for a significant post-starburst
population in many luminous AGNs, and that a direct, causal link may exist
between star formation and black hole accretion.  Quantifying the ongoing
star formation rate in AGNs, however, is much more challenging because of the
strong contamination by the active nucleus.
I discuss recent work attempting to measure the star formation rate
in luminous AGNs and quasars.  The exceptionally low level of coeval star
formation found in these otherwise gas-rich systems suggests that the star
formation efficiency in the host galaxies is suppressed in the presence of
strong AGN feedback.


\end{abstract}

\section{Introduction}           
As two of the most widely studied extragalactic phenomena, AGN and
starburst activity have often been conjectured to share a ``connection,''
although precisely what the nature of the connection is or how it comes about
is not always stated.  Many galaxies exhibit concurrent signatures of
both an AGN and a starburst, and much attention has been devoted to sorting
out which of the two processes dominates the energetics in these
hybrid systems.  The best-known example of such systems are the ultraluminous
infrared galaxies, whose energy source has been the subject of much intense
debate since their initial discovery over two decades ago (e.g., Soifer et al.
1984; Sanders \& Mirabel 1996; Genzel et al.  1998; Tacconi et al. 2002). Less
extreme examples of intermixed AGN and starburst activity have been frequently
reported in lower luminosity AGNs such as Seyfert galaxies (e.g., Boisson et
al. 2000; Gonz\'alez Delgado et al. 2001).  Implicit in many of
these studies, although often not always clearly articulated, is the assumption
that somehow the two processes involved---black hole accretion and star
formation---are causally linked.  Why must this be so?  Black holes are
nearly ubiquitous, at least in massive galaxies (see reviews in Ho 2004a), as
is AGN activity of varying levels of intensity (Ho 2004b).  And provided that
cold gas exists in galaxies, it has little choice but to form stars,
especially in the presence of dynamical perturbations.  When galaxies crash
together, as is the case in most ultraluminous infrared sources (e.g., Sanders
\& Mirabel 1996), why should it be surprising that we would find both
the AGN and the young stars to light up simultaneously?  What other option is
there, especially for galaxies selected from infrared or ultraviolet surveys?
To be sure, physically motivated evolutionary scenarios linking starbursts and
AGNs have been proposed (e.g., Sanders et al. 1988), but quantitatively
testing them is much trickier.  We must distinguish truly causal connections
from merely phenomenological ones.

There are a variety of concrete paths in which starbursts and AGNs might be
physically linked.  The collapse of very massive stars, especially in the
early Universe, may give rise to the first population of seed black holes.
According to some recent theories, compact super star clusters, which form
preferentially in starburst environments, may provide conditions particularly
conducive to the production of intermediate-mass black holes.
Once formed, black holes can be fed by stars, either directly through
tidal capture or indirectly through gas shed via stellar mass loss.
Likewise, the energy liberated by black hole accretion can either trigger
star formation (e.g., by dynamically compressing gas clouds through radio
jets) or suppress it (e.g., by blowing away all the gas through strong AGN
feedback).  More generally, the discovery of scaling relations between central
black hole masses and the bulge properties of their host galaxies (Magorrian
et al. 1998; Gebhardt et al.  2000; Ferrarese \& Merritt 2000) has stimulated
a plethora of ideas linking black hole growth with galaxy assembly.

le everyone agrees that the growth of black holes must be closely
linked with galaxy formation (e.g., Silk \& Rees 1998; Kauffmann \& Haehnelt
2000; Begelman \& Nath 2005), there is no consensus as
to exactly how accretion and star formation are really coupled.  Are they well
synchronized, or does one process precede the other, and if so, what is the
time lag?  When the black hole is actively growing, does the feedback from the
AGN actually facilitate or inhibit star formation in the host galaxy?
These important issues are unlikely to be settled through theoretical
speculations or numerical simulations alone.  Some empirical guidance from
observations would be highly desirable.

\section{Post-Starbursts in AGNs}

The host galaxies of AGNs of moderate to high luminosity frequently show
spectroscopic signatures of A-type stars indicative of an intermediate-age,
post-starburst stellar population (see review by
Heckman 2004).  The evidence has been most thoroughly documented in Seyfert
galaxies (e.g., Boisson et al. 2000; Gonz\'alez Delgado et al. 2001; Q. S. Gu
and M.  Imanishi, this meeting), but post-starbursts have also been reported in
bona fide quasars (Brotherton et al. 1999; Canalizo \& Stockton 2001;
Z. Shang, this meeting).  By contrast, the central (10--100 pc) regions of
weaker AGNs, such as low-luminosity Seyferts and LINERs, nearly always possess
an old stellar population (Ho et al. 2003; Sarzi et al. 2005).  Notwithstanding
the extensive evidence for intermediate-age stars in Seyferts and quasars, how
do we establish a truly causal connection between star formation and AGN
activity?  This crucial step was achieved by Kauffmann et al. (2003), whose
analysis of a large sample of narrow-line (Type 2) AGNs selected from the
Sloan Digital Sky Survey (SDSS) revealed not only the frequent presence of
intermediate-age stars, but, most importantly, that the post-starburst
fraction in these objects {\it increases}\ with increasing AGN luminosity.
This, in my view, constitutes the most convincing evidence to date for a
direct, statistically significant connection between AGN and star formation
activity.

As encouraging as this is, we must realize that the SDSS results of Kauffmann
et al. pertain only to the post-starburst phase, on timescales of
$\sim 10^8-10^9$ yr.  While the lifetime of AGNs is currently quite
uncertain, Martini (2004) estimates it to lie in the range of $10^6$ to
$10^8$ yr, precisely in the regime inaccessible by the current optical
absorption-line diagnostics.  In order to put meaningful constraints on
the time sequence of black hole and galaxy growth, we must attempt to estimate
the stellar population that is roughly coeval with the AGN, namely the
{\it ongoing}\ star formation rate (SFR) as imprinted by massive,
ionizing stars (ages \lax $10^7$ yr).
\section{Estimating Star Formation Rates in AGNs}

How can the current SFR in AGNs be estimated?  While a variety of SFR
estimators have been developed for normal (inactive) galaxies (e.g., Kennicutt
1998; Gilfanov et al. 2004), nearly all of them are problematic when applied to
active galaxies because of the strong confusion with emission from the AGN
itself.  For example, neither the mid-ultraviolet continuum strength nor the
recently proposed indicator using hard X-ray luminosity (e.g., Gilfanov et al.
2004) can be used, since both are ubiquitous in AGNs.  In detail the
ultraviolet and X-ray continuum properties of AGNs do differ from those of
starbursts, but most survey-quality data rarely have the luxury of discerning
this level of subtlety.  Most discussions in favor of one process versus the
other boil down to crude luminosity arguments based on precedence from
low-redshift observations.  The strength of the
radio synchrotron emission is often used as a star formation tracer, but it is
rendered useless in AGNs, even radio-quiet ones, because AGNs are never
totally radio-silent, and because we still lack a fully predictive theory to
explain the origin of jets in accretion-powered sources.  Some investigators
resort to morphological arguments: AGNs should be compact, whereas starbursts
should be extended.  I do not find such qualitative criteria very persuasive.
Clearly some AGNs do produce extended radio lobes, and many starbursts are
highly centrally concentrated.  The far-infrared (FIR) luminosity offers a
highly effective, reddening-insensitive measure of the SFR in inactive
galaxies, and it has even been used in this same capacity in quasars (e.g.,
Beelen
et al. 2004; P. Cox and X. Y. Xia, this meeting). However, we must worry the
extent to which we can truly separate AGN heating from stellar heating.  The
exact apportionment is model-sensitive, depending on the detailed geometrical
distribution of the dust.  It is often casually assumed that the cooler dust
component traced by the FIR continuum must be heated by stars, but, as
shown by Sanders et al. (1989), it can be equally modeled by AGN heating of
a warped disk.  Incidentally, the radio--FIR correlation offers
little clarification in this matter, since radio-quiet AGNs, apparently
fortuitously, exhibit radio/FIR ratios that are quite similar to those found in
starburst galaxies (see discussion in Sanders et al. 1989).
In terms of line emission, none of the standard hydrogen
recombination lines (e.g., H$\alpha$) offer a viable solution, since these
lines are extremely prominent in the AGN spectrum itself.

There is, however, one last recourse.

Spectroscopic
surveys of distant galaxies, particularly for redshifts between $z\approx0.4$
to 1, routinely use [O~{\sc II}] \lamb3727, a prominent nebular emission line in
\hii\ regions, to estimate SFRs.  Since its original introduction by Gallagher
et al. (1989), the utility of [O~{\sc II}] as a SFR indicator has been scrutinized by
numerous authors (e.g., Kennicutt 1998; Cardiel et al. 2003; Hopkins et al.
2003; Kewley et al. 2004).  The [O~{\sc II}] line suffers from two main drawbacks,
namely its sensitivity to dust extinction and metallicity effects.
Hopkins et al. (2003) find, from comparing SFRs derived from [O~{\sc II}] versus
SFRs derived from radio and FIR continuum, that while dust extinction
is certainly non-negligible, on average [O~{\sc II}] only suffers from an extinction
of $A_V \approx 1$ mag (corresponding to a factor of 4--5).  The amount of
dust extinction increases with increasing SFR, and the above estimate
applies to SFRs \lax\ 100 \solmass\ \peryr.  Kewley et al. (2004) have
evaluated the influence of metallicity variations, but this effect is
less serious than dust extinction; plausible metallicity uncertainties of a factor
of 2, for example, result in SFR variations of only $\sim$50\%.

Ho (2005) recently proposed that the [O~{\sc II}] line can be used as an equally
effective SFR estimator in the host galaxies of AGNs.  To mitigate confusion
by [O~{\sc II}] emission from the narrow-line region, this method should not be
applied to low-ionization AGNs (i.e. LINERs, which include many radio
galaxies), but rather should be limited to high-ionization sources, whose
intrinsic [O~{\sc II}] line is both observed and predicted to be weak.  In practice,
this requirement is not too restrictive, since virtually all relatively
luminous AGNs, including classical Seyferts and quasars, fall in this
category.  (The ionization state of AGNs generally correlates with their
luminosity.)  And, of course, it is the actively accreting sources that matter
most in terms of black hole growth.   If high-ionization AGNs experience
substantial levels of ongoing star formation, the integrated contribution from
H~{\sc II} regions will boost the strength of the [O~{\sc II}] line (compared to, say,
[O~{\sc III}] \lamb 5007, which can be largely ascribed to the AGN itself).
In any case, the observed [O~{\sc II}] strength, after applying reasonable corrections
for extinction and metallicity, provides an absolute upper limit to the
total ongoing SFR in the host galaxy.

\section{Recent Results}

Using the [O~{\sc II}] technique described above, Ho (2005) reached some surprising
conclusions, which I will summarize here, regarding the ongoing level of star
formation in quasars.  First and foremost, the [O~{\sc II}] line is {\it always}\
very weak.  Although measurements of the [O~{\sc II}] line in individual quasars
are not widely published, statistical averages of very large samples of
objects, as captured in ``composite spectra,'' are readily available
for nearly all of the extant major quasar surveys (LBQS, FIRST, 2dF+6dF,
SDSS).  Examination of these composite spectra reveals a common trend:
[O~{\sc II}] is clearly detected, but it is very weak, generally having
a strength equal to 10\%--20\% of that of [O~{\sc III}] \lamb5007.  This relative
intensity is entirely consistent with a pure AGN origin, leaving little
additional room for a significant contribution from \hii\ regions.  As an
illustrative example, the LBQS composite of Francis et al. (1991) gives an
[O~{\sc II}] equivalent width of 1.9 \AA.  For an average quasar luminosity of
$\langle M_B \rangle \approx -23.5$ mag, an average redshift of
$\langle z \rangle \approx 1.3$,
and a continuum spectrum $f_\nu \propto \nu^{-0.32}$ (Francis et al. 1991),
we find
$\langle L_{\rm [O~{\sc II}]} \rangle = 9.8\times 10^{41}$
\lum.\footnote{The following cosmological parameters are adopted: $H_0$ =
72 \kms~Mpc$^{-1}$, $\Omega_{\rm m} = 0.3$, and $\Omega_{\Lambda} = 0.7$.}  To
convert this line luminosity into a SFR, we use the calibration of Kewley
et al. (2004), adopting three simple assumptions (see Ho 2005 for a more
detailed justification): (1) that the amount of dust extinction in AGN host
galaxies is comparable to that deduced for moderately actively star-forming
galaxies ($A_V \approx 1$ mag; e.g., Hopkins et al. 2003); (2) that the
metallicity is twice solar (e.g., Storchi-Bergmann et al. 1998); and (3) that
one-third of the observed [O~{\sc II}] strength comes from \hii\ regions (the rest
attributed to the AGN).  These assumptions lead to
$\langle {\rm SFR} \rangle \approx 7$ \solmass\ \peryr.

\vskip 0.3cm

\begin{figure}
\hskip +1.2in
\psfig{file=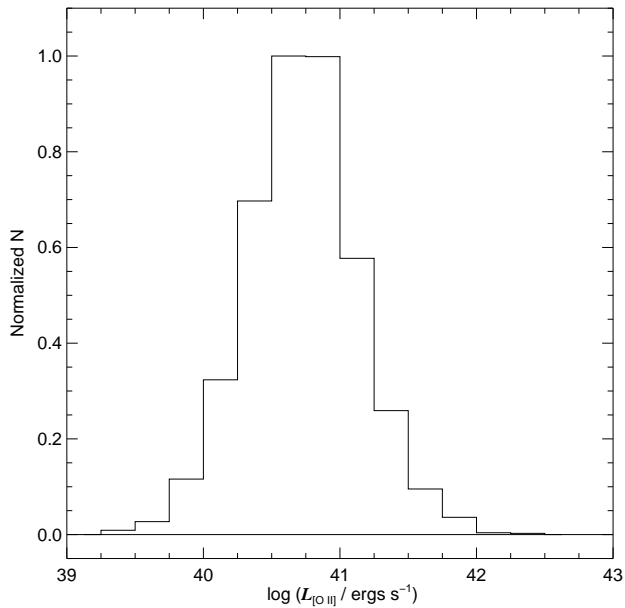,width=8.5cm,angle=0}
\caption{
The distribution of [O~{\sc II}] \lamb3727 luminosities for a sample of
$\sim 3600$ Type 1 AGNs selected from SDSS.  No extinction correction
has been applied because the observed Balmer decrements indicate very
little internal extinction.  Adapted from Kim, Ho, \& Im (2005).
}
\end{figure}
\vskip 0.6cm

\begin{figure*} [t]
\vskip +10mm
\centerline{\psfig{file=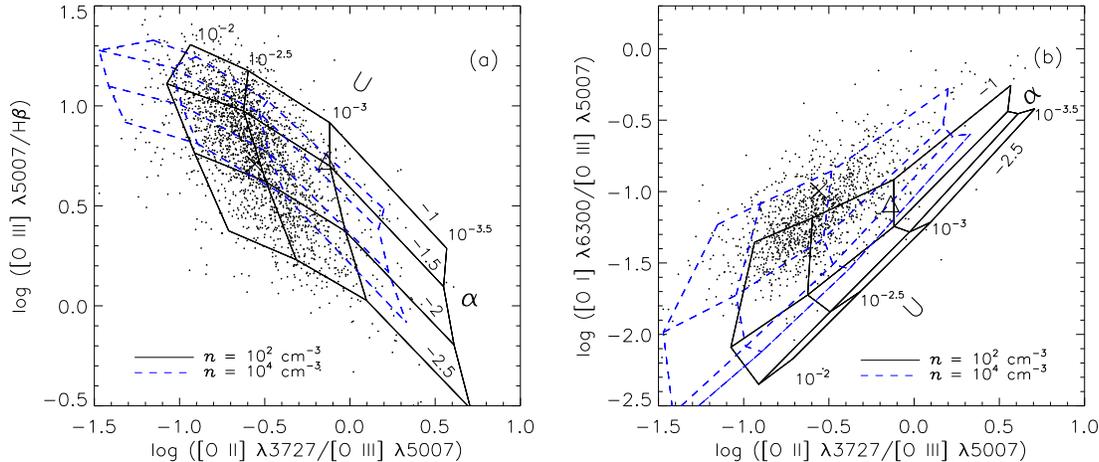,width=15.0cm,angle=0}}
\caption{
Line ratios of Type 1 AGNs (small dots) and Type 2 quasars (large triangle,
from composite spectrum of Zakamska et al. 2003, with internal extinction
correction applied) compared with photoionization models.  The large cross
marks the location of the average value of the upper limits in our sample,
which are consistent with the detections.  Diagnostic diagrams showing
(\textit{a}) \oiii\ \lamb5007$/$\hbet\ vs.  \oii\ \lamb3727$/$\oiii\
\lamb5007 and (\textit{b}) \oi\ \lamb6300$/$\oiii\ \lamb5007 vs.  \oii\
\lamb3727$/$\oiii\ \lamb5007.  Two grids are shown, representing hydrogen
densities $n=10^2\,{\rm cm}^{-3}$ ({\it solid line}) and
$n=10^4\,{\rm cm}^{-3}$ ({\it dashed line}).  Each grid shows models for
an ionization parameter of $U = 10^{-2}$, $10^{-2.5}$, $10^{-3}$, and,
$10^{-3.5}$ ({\it left to right}) and spectral index for the ionizing
continuum of $\alpha = -2.5$, $-2$, $-1.5$, and $-1$ ({\it bottom to top}).
See Kim, Ho, \& Im (2005) for details.
}
\end{figure*}
\vskip +5mm

As this result relied on composite spectra, Ho also examined a more limited
sample of 25 nearby Palomar-Green (PG) quasars that have available individual
measurements of [O~{\sc II}] flux.  In this case, the limits on the current
level of star formation are even more stringent.  On average the [O~{\sc II}]
line strengths (or their upper limits) translate into a SFR of $\sim 1$
\solmass\ \peryr.  To place this result on a firmer statistical footing,
Kim, Ho, \& Im (2005) have performed a systematic analysis of the narrow
emission-line spectra of a homogeneous, statistically complete sample of
$\sim$3600 Type 1 AGNs, selected to have redshifts $< 0.3$ from the Third Data
Release of SDSS.  The [O~{\sc II}]  luminosities are quite modest (Fig. 1), ranging
from $10^{40}$ to $10^{42}$ ergs~s$^{-1}$, with an average of
$\langle L_{\rm [O~{\sc II}]} \rangle = 5.2 \times 10^{40}$ \lum. Importantly,
the observed [O~{\sc II}] line strength, when viewed in the context of the rest of
the optical narrow-line spectrum, is again entirely consistent with a
pure AGN origin.  This is shown in Figure 2, where the observed line ratios
are compared to a new set of photoionization models.  The measured range of
\oii/\oiii\ ratios can be easily reproduced by assuming relatively standard
parameters for the AGN narrow-line region.  In other words, there is no
need for any additional source of excitation, such as that coming from
hot, young stars.  Nevertheless, adopting, as before, the conservative
assumption that one-third of the \oii\ strength comes from star formation, we
arrive at a mildly startling conclusion:
$\langle {\rm SFR} \rangle \approx 0.5$ \solmass\ \peryr.  To better
appreciate the peculiarity of this result, it is instructive to recall
that the integrated SFRs of local spiral galaxies, including the Milky Way,
lie in the range of $\sim 1-3$ \solmass\ \peryr\ (Solomon \& Sage 1988;
Scoville \& Good 1989).

Could the SFRs, especially of the PG and SDSS AGNs, be exceptionally low
because the host galaxies are early-type (e.g., S0 and E) systems?  Dunlop
et al. (2003) conclude that low-redshift quasars predominantly reside in
massive, evolved, early-type galaxies.  If all low-redshift quasars have
early-type, gas-poor hosts, this would explain why their SFRs are so low.
Unfortunately, the existing imaging data for the SDSS sample do not provide
strong constraints on the detailed morphologies of the host galaxies. However,
the situation is much better for the PG sample, a number of which have
been imaged at high resolution with the {\it Hubble Space Telescope}\
(Bahcall et al. 1997; Barth et al. 2004; Veilleux et al. 2005).  With few
exceptions, the host galaxies of PG quasars generally tend to exhibit a
fairly prominent disk component, often accompanied by visible spiral arms
and tidal features.  The host galaxies of PG quasars are not exclusively
giant ellipticals, not by a long shot.  At least from a morphological point of
view, many of them resemble bulge-dominated disk galaxies, and, as such,
their SFRs should be comparable to, or, naively, perhaps even greater than,
those of inactive, normal spirals.  This appears not to be the case.

\vskip -0mm
\begin{figure*}
\centerline{\psfig{file=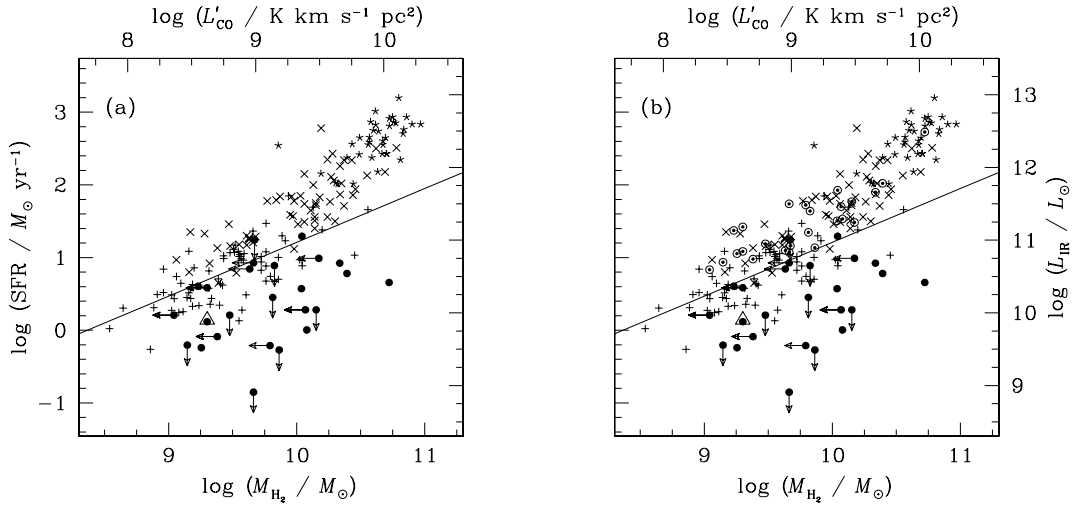,width=15.0cm,angle=270}}
\caption{
({\it a}) The dependence of SFR on molecular gas content in galaxies; the
right and top axes give the alternative representation in terms of infrared
luminosity and CO luminosity, respectively.  The PG quasars from Ho (2005)
are plotted as {\it filled circles}, with their SFRs luminosities
estimated from \oii\ measurements; upper limits are denoted with arrows.
Three galaxy samples are included for
comparison: isolated and weakly interacting galaxies ({\it pluses}, with
best-fitting line (Solomon \& Sage 1988), luminous infrared galaxies
({\it crosses}), and ultraluminous infrared galaxies ({\it asterisks}).
The large {\it triangle}\ marks the location of the Milky Way.
({\it b}) Same as ({\it a}), but with the total infrared luminosities
actually observed in the PG quasars marked with {\it semi-filled circles}.
See Ho (2005) for details.
}
\end{figure*}
\vskip 0.3cm

Suppose the host galaxies, despite having a prominent disk component and
spiral arms, somehow possess less gas than usual, as has been suggested
for ``anemic'' spirals (e.g., Bothun \& Sullivan 1980).  Or perhaps most of
the gas has been blown out of the galaxy as a result of very efficient AGN
feedback, as suggested by some numerical simulations.  This, too, appears not
to be so.  In a recent millimeter interferometric survey of a complete
subsample of the nearest PG quasars, Scoville et al. (2003) find that the
majority of them in fact contain significant amounts of CO emission.  Most of
the CO emission remains unresolved within a 4\asec\ beam, implying that the
gas is confined within the central $\sim 5$ kpc of the galaxies.  This
result is noteworthy in that Scoville et al. deliberately avoided any
infrared selection that might bias the sample toward unusually dusty objects.
Assuming a standard (Galactic) CO-to-H$_2$ conversion factor, the implied
molecular gas content ranges from $\sim 10^9$ to $10^{10}$ \solmass. Again, for
comparison, the Milky Way is characterized by $M_{\rm H_2} \approx2\times10^9$
\solmass.  The CO survey of Evans et al. (2001) further corroborates the
view that PG quasars tend to be gas-rich (although the infrared selection
applied in this study complicates the interpretation), as does the nearly
ubiquitous detection of large quantities of dust by Haas et al. (2003) using
{\it ISO}.

Now we are faced with an even deeper dilemma: Why are the SFRs so low,
{\it in spite}\ of there being plenty of gas?  To emphasize this point,
Figure 3{\it a}\ shows the SFR plotted versus the molecular gas mass.  In this
star formation ``efficiency'' diagram, regular galaxies follow a roughly
defined locus delineated by the diagnonal solid line; infrared-luminous
galaxies lie somewhat offset above the line; and ultraluminous infrared
galaxies lie even more displaced still, illustrating the well-known fact that
the most extreme starbursts convert gas to stars at a much higher efficiency.
Within this landscape, it is quite striking that the PG quasars studied
by Ho (2005) fall significantly below the locus not only of starburst
galaxies, but also that of normal galaxies.

The interpretation of the above results, of course, is not immune from
criticism.  A number of uncertainties, such as extinction, may affect the
SFRs estimated from the \oii\ emission line, and the applicability of the
Galactic CO-to-H$_2$ conversion factor may be debated.  These concerns
have been discussed in Ho (2005), and will not be repeated here. Suffice it
to say, reasonable assumptions have been adopted, but currently it is
difficult to prove whether these assumptions are correct or not.  If they
turn out to be wrong (e.g., the star-forming regions in AGN host galaxies
are pathologically much more extincted than inactive galaxies of similar
type and gas-richness), then we would be faced with a new challenge of
having to explain them.   Taken at face value, the current evidence indicates
that {\it the presence of a luminous AGN suppresses the star formation
efficiency within the host galaxy}.

\section{Origin of the Far-Infrared Emission in Quasars}

While most people now agree that the FIR continuum in AGNs arises from thermal
dust emission, there is no universal concensus on the origin of the dust's
heating source (e.g., Haas et al. 2003).  Since the FIR continuum traces
relatively cool dust temperatures, it is often taken for granted that the
dust must be heated by stars, and many investigators proceed to utilize the
FIR luminosity to derive SFRs, neglecting any possible contribution from the
AGN.  Enormous SFRs have been inferred in some
high-redshift quasars, especially when molecular gas has been detected
(e.g., Beelen et al. 2004).  But as mentioned in \S~3, the cooler dust
component traced by the FIR does not automatically dictate that stars dominate
the energetics; an extended, warped disk illuminated by a central AGN can
easily mimic this signature (e.g., Sanders et al. 1989).

The \oii-derived SFRs discussed above provide a new, independent constraint on
the origin of the FIR emission in quasars, and it serves as a warning that the
mere existence of large quantities of molecular gas in infrared-luminous
systems, in the presence of a luminous AGN, does not guarantee that stars form
at a high rate.  If we were to blindly attribute all of the infrared
emission to stellar heating, the SFRs deduced for the PG sample would
be at least an order of magnitude larger than those obtained from the \oii\
line (Fig. 3{\it b}; see Table 1 in Ho 2005).  Taking the \oii-derived
SFRs at face value, it appears that the bulk of the infrared emission in
the (optically selected) PG quasars is powered by accretion energy rather than
young stars.

\section{Type 2 Quasars, A Different Kind of Beast?}

By contrast to normal (Type 1) quasars, which have low SFRs, the level of
star formation activity in Type 2 quasars appears to be significantly more
elevated.  This somewhat unexpected result is discussed by Kim, Ho, \& Im
(2005), who find, from inspection of the composite spectrum of SDSS Type 2
quasars published by Zakamska et al. (2003), that the \oii\ line in Type 2
quasars is approximately an order of magnitude stronger (relative to \oiii,
for the same \oiii\ luminosity) than in Type 1 quasars.  The inferred SFR is
$\sim 20$ \solmass\ \peryr, equivalent to that of a modest local starburst.
Apart from the obvious implication that Type 1 and Type 2 quasars are
not intrinsically the same (modulo viewing angle), we might postulate that
the source of obscuration in the Type 2 sources arises not from a standard
compact torus, but perhaps instead from more patchy or extended dust
associated with the more abundant star-forming regions in the host galaxy.
Without too much of a stretch, we can further speculate that perhaps
Type 2 quasars in fact are the {\it precursors}\ of Type 1 quasars.

\section{AGN Feedback}

Our appraisal of the strength of the \oii\ emission line in Type 1 AGNs, based
either on statistical averages of large number of sources or measurements of
individual sources, leads to the conclusion that the ongoing SFR in the host
galaxies is very modest.  The low SFRs stem not from a deficiency in gas,
for abundant molecular gas has been detected, at least for a representative
sample of nearby sources.  Evidently gas can withstand the ravages of AGN
feedback, presumably because the gas preferentially lies in a plane, which,
under most circumstances, has a small cross-section with the intrinsically
anisotropic feedback field (either radiation or jets) of the
AGN. The real mystery is why the gas, despite being plentiful,
fails to form stars when the AGN is simultaneously most active (when it is
viewed as a full-blown, optically revealed Type 1 AGN or quasar).  What could
be responsible for the anomalously low star formation efficiencies?  One can
easily envisage a number of ways in which the hard radiation field of an AGN
can have a profound impact on the thermal and ionization structure of a
molecular cloud (e.g., Maloney 1999; (Begelman 2004; Di~Matteo et al. 2005),
but precisely how this leads to suppression of star formation remains to be
fully elucidated.  Clearly, simple prescriptions for star formation, such as
the Schmidt law, usually adopted in numerical simulations of galaxy formation
need to be evaluated more judiciously.  I hope that theorists would take up
the challenge to investigate this problem.

\begin{acknowledgements}
This work was funded by the Carnegie Institution of Washington and by NASA
grants from the Space Telescope Science Institute (operated by AURA, Inc.,
under NASA contract NAS5-26555).  I thank the organizers for partial financial
support, and for holding the meeting in unforgetable Lijiang.  Some of the
results presented here were obtained in collaboration with M. Kim and M. Im.
\end{acknowledgements}

\label{lastpage}


\begin{thebibliography}
\baselineskip=12pt
\parskip=-3pt


\bibitem[]{}
Bahcall, J.~N., Kirhakos, S., Saxe, D.~H.,  Schneider, D.~P., 1997, \apj,
479, 642

\bibitem[]{}
Barth, A.~J., Nelson, C. H., Martini, P.,  Ho, L.~C., 2004, BAAS, 205, 143.09

\bibitem[]{}
Beelen, A., et al., 2004, \aap, 423, 441

\bibitem[]{}
Begelman, M.~C., 2004, In: ed. L. C. Ho, ed., Carnegie Observatories Astrophysics Series, Vol. 1: Coevolution of Black Holes and Galaxies, Cambridge: Cambridge
Univ. Press, p. 375

\bibitem[]{}
Begelman, M.~C.,  Nath, B. B., 2005, \mnras, 361, 1387

\bibitem[]{}
Boisson, C., Joly, M., Moultaka, J., Pelat, D.,  Serote Roos, M., 2000,
\aap, 357, 850

\bibitem[]{}
Bothun, G.~D.,  Sullivan, W.~T., III, 1980, \apj, 242, 903

\bibitem[]{}
Brotherton, M.~S., et al., 1999, \apj, 520, L87

\bibitem[]{}
Brotherton, M.~S., Grabelsky, M., Canalizo, G., van Breugel, W., Filippenko, 
A.~V., Croom, S., Boyle, B.,  Shanks, T., 2002, \pasp, 114, 593

\bibitem[]{}
Brotherton, M.~S., Tran, H.~D., Becker, R.~H., Gregg, M.~D., 
Laurent-Muehleisen, S.~A.,  White, R.~L., 2001, \apj, 546, 775

\bibitem[]{}
Canalizo, G.,  Stockton, A., 2001, \apj, 555, 719

\bibitem[]{}
Cardiel, N., Elbaz, D., Schiavon, R.~P., Willmer, C.~N.~A., Koo, D.~C.,  
Phillips, A.~C.,  Gallego, J., 2003, \apj, 584, 76

\bibitem[]{}
Di Matteo, T., Springel, V.,  Hernquist, L., 2005, Nature, 433, 604

\bibitem[]{}
Dunlop, J.~S., McLure, R.~J., Kukula, M.~J., Baum, S.~A., O'Dea, C.~P.,
 Hughes, D.~H., 2003, \mnras, 340, 1095

\bibitem[]{}
Evans, A.~S., Frayer, D.~T., Surace, J.~A.,  Sanders, D.~B., 2001, \aj, 
121, 1893

\bibitem[]{}
Ferrarese, L.,  Merritt, D., 2000, \apj, 539, L9

\bibitem[]{}
Francis, P.~J., Hewett, P.~C., Foltz, C.~B., Chaffee, F.~H., Weymann, R.~J.,
 Morris, S.~L., 1991, \apj, 373, 465

\bibitem[]{}
Gallagher, J.~S., Bushouse, H.,  Hunter, D.~A., 1989, \aj, 97, 700

\bibitem[]{}
Gebhardt, K., et al., 2000, \apj, 539, L13

\bibitem[]{}
Genzel, R., et al., 1998, \apj, 498, 579

\bibitem[]{}
Gilfanov, M., Grimm, H.-J.,  Sunyaev, R., 2004, \mnras, 347, L57

\bibitem[]{}
Gonz\'alez Delgado, R.~M., Heckman, T.,  Leitherer, C., 2001, \apj, 546, 845

\bibitem[]{}
Haas, M., et al., 2003, \aap, 402, 87

\bibitem[]{}
Heckman, T. M., 2004, In: ed. L. C. Ho, ed., Carnegie Observatories Astrophysics Series, Vol. 1: Coevolution of Black Holes and Galaxies, Cambridge: Cambridge
Univ. Press, p. 359

\bibitem[]{}
Ho, L.~C., 2004a, ed., Carnegie Observatories Astrophysics Series, Vol. 1:
Coevolution of Black Holes and Galaxies (Cambridge: Cambridge Univ. Press)

\bibitem[]{}
------., 2004b, In: ed. L. C. Ho, ed., Carnegie Observatories Astrophysics Series, Vol. 1: Coevolution of Black Holes and Galaxies, Cambridge: Cambridge
Univ. Press, p. 293

\bibitem[]{}
------., 2005, ApJ, 629, 680

\bibitem[]{}
Ho, L.~C., Filippenko, A.~V.,  Sargent, W.~L.~W., 2003, \apj, 583, 159

\bibitem[]{}
Hopkins, A.~M., et al., 2003, \apj, 599, 971

\bibitem[]{}
Kauffmann, G., et al., 2003, \mnras, 346, 1055

\bibitem[]{}
Kauffmann, G.,  Haehnelt, M. G., 2000, \mnras, 311, 576

\bibitem[]{}
Kennicutt, R.~C., 1998, \annrev, 36, 189

\bibitem[]{}
Kewley, L.~J., Geller, M.~J.,  Jansen, R.~A., 2004, \aj, 127, 2002

\bibitem[]{}
Kim, M., Ho, L. C.,  Im, M., 2005, \apj, submitted

\bibitem[]{}
Magorrian, J., et al., 1998, \aj, 115, 2285

\bibitem[]{}
Maloney, P. R., 1999, ApSS, 266, 207

\bibitem[]{}
Martini, P., 2004, In: ed. L. C. Ho, ed., Carnegie Observatories Astrophysics Series, Vol. 1: Coevolution of Black Holes and Galaxies, Cambridge: Cambridge
Univ. Press, p. 170

\bibitem[]{}
Sanders, D.~B.,  Mirabel, I.~F., 1996, \annrev, 34, 749 

\bibitem[]{}
Sanders, D.~B., Phinney, E.~S., Neugebauer, G., Soifer, B.~T.,  Matthews, K. 
1989, \apj, 347, 29

\bibitem[]{}
Sanders, D.~B., Soifer, B.~T., Elias, J.~H., Madore, B.~F., Matthews, K.,
Neugebauer, G.,  Scoville, N.~Z., 1988, \apj, 325, 74

\bibitem[]{}
Sarzi, M., Rix, H.-W., Shields, J.~C., Ho, L.~C., Barth, A. J., Rudnick,
G., Filippenko, A.~V.,  Sargent, W.~L.~W., 2005, \apj, 628, 169

\bibitem[]{}
Scoville, N.~Z., Frayer, D.~T., Schinnerer, E.,  Christopher, M., 2003,
\apj, 585, L105

\bibitem[]{}
Scoville, N.~Z.,  Good, J. C., 1989, \apj, 339, 149

\bibitem[]{}
Silk, J.,  Rees, M. J., 1998, \aap, 331, L1

\bibitem[]{}
Soifer, B.~T., et al., 1984, \apj, 278, L71

\bibitem[]{}
Solomon, P.~M.,  Sage, L.~J., 1988, \apj, 334, 613

\bibitem[]{}
Storchi-Bergmann, T., Schmitt, H.~R., Calzetti, D.,  Kinney, A.~L., 1998,
\aj, 115, 909

\bibitem[]{}
Tacconi, L.~J., Genzel, R., Lutz, D., Rigopoulou, D., Baker, A.~J.,
Iserlohe , C.,  Tecza, M., 2002, \apj, 580, 73

\bibitem[]{}
Veilleux, S., Kim, D. C., Peng, C. Y., Ho, L. C., Tacconi, L. J.,
Genzel, R., Lutz, D.,  Sanders, D. B., 2005, \apj, submitted

\bibitem[]{}
Zakamska, N.~L., et al., 2003, \aj, 126, 2125
\end{thebibliography}
\end{document}